\documentclass[12pt]{article}
\usepackage{epsfig,graphics}
\newcommand{\be}{\begin{equation}}
\newcommand{\bea}{\begin{eqnarray}}
\newcommand{\eea}{\end{eqnarray}}
\newcommand{\ba}{\begin{array}}
\newcommand{\ea}{\end{array}}
\newcommand{\ee}{\end{equation}}

\expandafter\ifx\csname mathbbm\endcsname\relax

\else

\fi \textheight 22cm \textwidth 15cm \topmargin 1mm \oddsidemargin
5mm \evensidemargin 5mm

\begin{document}
\begin{titlepage}
\hfill \vbox{
    \halign{#\hfil         \cr
           } 
      }  
\vspace*{20mm}
\begin{center}
{\Large {\bf  ON Dielectric Membranes}
\\}

\vspace*{15mm} \vspace*{1mm} {Mohammad A. Ganjali}
 \\
\vspace*{1cm}

{Department of Fundamental Sciences, Tarbiat Moaallem University\\
Tehran, Iran\footnote {Ganjali@theory.ipm.ac.ir}}

 \vspace*{1cm}
\end{center}
\begin{abstract}
 By placing a collection of membranes in 11 dimensional background supergravity six form
 field and considering the non-linear action of multiple membranes
 we obtain the potential for transverse scalar fields of membranes at leading order and
 study some possible vacuum configurations. We find
 that the system has a stable vacuum in which is a formation of
 membranes into fuzzy three sphere.
\end{abstract}

\end{titlepage}

\section{Introduction}
When a neutral material put in an external electric field the
electric charges separate from each other and form an electric dipole.
 Similar phenomena exists in string theory where there are
extended objects such as $D_p$ branes and higher rank antisymmetric
form fields as charges of these objects. In particular, When a set
of neutral $D$ branes put in an external antisymmetric background
field one observes the "polarized" $D$ branes which are expanded
into a higher dimensional world volume theory \cite{Myers:1999ps}.

One may naturally expect such phenomena for membranes in M-theory.
There are some efforts to studying  this in the context of
M-theory \cite{Arai:2008kv} where it was used some idea such as
considering massive ABJM theory \cite{Aharony:2008ug} or
Basu-Harvey equation \cite{Basu:2004ed} as boundary conditions of
membranes. But recently, some wonderful works was done in
understanding the dynamics of multiple $M2$ branes. Firstly, it
was found an action \cite{Bagger:2006sk} which was based on an
strange algebra, Lie-3 algebra, which is expected to describe the
low energy dynamics of multiple membranes. Based on this action,
the authors of \cite{Iengo:2008cq} found a non-linear action for
$M2$ branes which in low energy limit reduces to that of
\cite{Bagger:2006sk}. Beside, the Lie-3 algebra valued fields
allows one to write down an action for the coupling of
supergravity background form fields $C_3$ and $C_6$ to world
volume of membranes \cite{Li:2008ez}.

 So, by using the nonlinear
action of multiple membranes and the action of couplings of
background form fields to world volume of membranes one can study
the effects of placing multiple membranes in these external
fields. The aim of this note is to study it in details.

 In particular, we will find that when a collection of membranes
are is placed in a constant background $C_6$ fields the
configuration of membranes has a trivial vacua in which the
potential $V(X)=0$ and all fields are represented by some
commuting matrices. But, the theory still has a nontrivial vacua
in which has lower energy and so is stable in comparison with
commuting one.
 Such vacuum is a configuration of membranes which are formed in fuzzy
 three sphere $S^3$ and are ended on five branes.
\section{On Triple Product}
First of all, we briefly introduce
triple product\footnote{Note that the main
results of this paper can be obtained without the assumption of
triple algebra on the theory. }. All fields in BLG theory are
valued in a Lie-3 algebra and can be expanded in terms of
generators of the algebra as\footnote{In this paper we use the
indices as $a,b,...=+,-,1,2,...,Dim \;{\cal G}$;
$\mu,\nu,...=0,1,2$; $\hat{I},\hat{J},...=0,1,...,10$ and $I,J,...=3,4,...,
10$.}\footnote{And similarly, one can expand the gauge fields and
spinor fields in terms of the generators.} $X^{I}=X^I_aT^a$
 in which the generators $T^a$
construct the following product
 \bea\label{m1}
 [T^a,T^b,T^c]=f^{abc}_{\hspace{.5cm}d}T^d,
  \eea
 where $f^{abc}_{\hspace{.5cm}d}$ are some structure constants and
 obey a fundamental identity
\bea \label{h6}
 f^{cde}_{\hspace{.5cm}g}f^{abg}_{\hspace{.5cm}h}=
f^{abc}_{\hspace{.5cm}g}f^{gde}_{\hspace{.5cm}h}
+f^{abd}_{\hspace{.5cm}g}f^{gec}_{\hspace{.5cm}h}+
f^{abe}_{\hspace{.5cm}g}f^{cdg}_{\hspace{.5cm}h}, \eea
It has
been done a lot of efforts to finding solutions of (\ref{h6}), see for
example
\cite{Bagger:2006sk,Li:2008ez,Ho:2008ei,Benvenuti:2008bt,Gomis:2008uv,Papadopoulos:2008sk}.
 An interesting solution was presented in
 \cite{Ho:2008ei,Benvenuti:2008bt,Gomis:2008uv}
 in which by decomposing $T^a=\{T^-,T^+,T^a\}$ one may demand
 \bea\label{h2}
[T^-,T^+,T^a]=0,\;\;&&\;\; [T^a,T^b,T^c]=f^{abc}T^-\cr
[T^+,T^a,T^b]&=&[T^a,T^b]=f^{ab}_{\hspace{.3cm}c}T^c
 \eea
 In this fashion, the $T^-$ mode completely decouples
 from the theory and one obtains a theory with
 ordinary Lie algebra generators $T^a$ which is lifted by
 $T^+$.
\\\\
\section{Non-linear action for $M2$ branes}
Proposed bosonic part of non-linear Lagrangian for
multiple membranes for gauge group $U(N)$ is given by
 \bea\label{f1}
 {\cal L}_{M_2}&=&-T_2STr\left
((det(S))^{1/4}\sqrt{-det\left(\eta_{\mu\nu}+
\frac{1}{T_2}\tilde{D}_{\mu}X^I\tilde{R}^{IJ}\tilde{D}_{\nu}X^J\right)}\right)\\
&&+\frac{1}{2}STr\left(\epsilon^{\mu\nu\lambda}\left(B_{\mu}+\frac{\partial_{\mu}X_+.X-X_
+.D_{\mu}X}{X_+^2}\right)\left(F_{\nu\lambda}+\frac{1}{T_2}
\tilde{D}_{\mu}X^I\tilde{P}^{IJ}\tilde{D}_{\nu}X^J \right)\right)\cr
&&+(\partial_{\mu}X_-^I-Tr(B_{\mu}X^I))\partial^{\mu}X_+^I-Tr\left(\frac{X_+.X}
{X_+^2}\hat{D}_{\mu}X^I\partial^{\mu}X_+^I-\frac{1}{2}
\left(\frac{X_+.X}{X_+^2}\right)^2\partial_{\mu}X_+^I\partial^{\mu}X_+^I\right)\nonumber
\eea
 where $A_{\mu},B_{\mu},X^I$ are in adjoint representation of $U(N)$
and $X^I_-,X^I_+$ are singlet under $U(N)$, and \bea\label{f2}
M^{IJK}&\equiv&X^I_+[X^J,X^K]+X^J_+[X^K,X^I]+X^K_+[X^I,X^J]\cr
\hat{D}_{\mu}X^I&=&D_{\mu}X^I-X^I_+B_{\mu}\,,\qquad
D_{\mu}X^I=\partial_{\mu}X^I+i[A_{\mu},X^I],
 \eea
  The above Lagrangian  is
invariant under global $SO(8)$ transformation and under $U(N)$
gauge transformation associated with the $A_{\mu}$. There is also
another transformation associated with the $B_{\mu}$ gauge field
which leaves the Lagrangian invariant
 \bea\label{f3}
 &&\delta_BX^I=X^I_+\Lambda\,,\;\;\;\;\;\;\delta_B B_{\mu}=D_{\mu}\Lambda\,,\cr
&& \delta_BX^I_+=0\,,\;\;\;\;\;\;\;\;\; \delta _BX^I_-=Tr(X^I\Lambda)
 \eea
 It is
important to note that the equation of motion for $X_-^I$ gives
$\partial_{\mu}\partial^{\mu}X_+^I=0$. Gauging the shift symmetry
$X_-^I\rightarrow X_-^I+C^I$ as \cite{Bandres:2008kj} by
introducing a new field $C_{\mu}^I$ and rewriting $\partial_{\mu}X_-^I$ as
$\partial_{\mu}X_-^I-C_{\mu}^I$, then the equations of motion of the new
fields give $\partial_{\mu} X_+^I=0$ which means that $X_+^I=v^I$ for
some constant $v^I$.
\\\\
\section{Membranes coupled to fluxes}

The coupling of the
antisymmetric fields $C_3$ and $C_6$ to world volume of $M2$
branes was given in \cite{Li:2008ez} as
 \bea\label{f4} {\cal
L}_{MCS}&=&\lambda_1 \epsilon^{\lambda\mu\nu}C_{IJK}STr(T^aT^bT^c)
D_{\lambda}X_a^ID_{\mu}X^J_bD_{\nu}X_c^K\hspace{.5cm}\\
&+&\lambda_2
\epsilon^{\lambda\mu\nu}C_{IJKLMN}STr([T^d,T^e,T^f]T^aT^bT^c)X_d^IX_e^JX_f^K
D_{\lambda}X_a^LD_{\mu}X^M_bD_{\nu}X_c^N\nonumber
 \eea
 Following \cite{Myers:1999ps}, and assuming all fields are valued in a
\textsl{non-associative algebra}, the above  Lagrangian may be
written in a generalized form as
 \bea\label{f5}
S_{MCS}=\mu_2\int{\left(P\left(e^{i\lambda
<\textbf{i}_{X}\textbf{i}_{X}\textbf{i}_{X}>}\Sigma
C_{(n)}\right)\right)}
 \eea
 where $\textbf{i}_{X}$ denotes the
interior product by $X^{\hat{I}}$ as a vector in transverse space and we define the operator $<\textbf{i}_{X}\textbf{i}_{X}\textbf{i}_{X}> $ as
\bea
 <\textbf{i}_{X}\textbf{i}_{X}\textbf{i}_{X}>=<X^{\hat{I}},X^{\hat{J}},X^{\hat{K}}>,\nonumber
\eea
and the associators is defined as \cite{Bagger:2006sk}
 \bea\label{f6}
<X^{\hat{I}},X^{\hat{J}},X^{\hat{K}}>=(X^{\hat{I}}.X^{\hat{J}}).X^{\hat{K}}-X^{\hat{I}}.(X^{\hat{J}}.X^{\hat{K}}),
 \eea
 The $\lambda$ is
a constant with dimension $\frac{1}{length^3}$. Defining the
triple product as
 \bea
[X^{\hat{I}},X^{\hat{J}},X^{\hat{K}}]&=&<X^{\hat{I}},X^{\hat{J}},X^{\hat{K}}>+<X^{\hat{J}},X^{\hat{K}},X^{\hat{I}}>
+<X^{\hat{K}},X^{\hat{I}},X^{\hat{J}}>\cr
&-&<X^{\hat{I}},X^{\hat{K}},X^{\hat{J}}>-<X^{\hat{J}},X^{\hat{I}},X^{\hat{K}}>-<X^{\hat{K}},X^{\hat{J}},X^{\hat{I}}>,
 \eea and expanding
(\ref{f5}) in power of $\lambda$ one finds
 \bea\label{f7}
S_{MCS}&=&\mu_2\int{STr\left(P\left(\frac{1}{6}C_{{\hat{I}}{\hat{J}}{\hat{K}}}DX^{\hat{I}}\wedge DX^{\hat{J}}\wedge DX^{\hat{K}}\right)\right)}\\
&+&i\lambda\mu_2\int STr\left(P\left(\frac{1}{216}[X^{\hat{I}},X^{\hat{J}},X^{\hat{K}}]
C_{{\hat{I}}{\hat{J}}{\hat{K}}{\hat{L}}{\hat{M}}{\hat{N}}}DX^{\hat{L}}\wedge DX^{\hat{M}}\wedge DX^{\hat{N}}\right)\right).\nonumber
 \eea
Obviously for $I,J={\hat{I}},{\hat{J}}=3,4,...,9$ one can reproduce (\ref{f4}).
 Note
that, although we define (\ref{f5}) by assuming the
non-associativity of fields, one still can write the action
(\ref{f7}) by demanding that all fields are valued in a triple
product algebra only without using the non-associativity.
\\\\
\section{Dielectric membranes}
First of all, recalling (\ref{h2})
and using the gauge symmetry of the BLG action
\cite{Bandres:2008kj}, to have the gauge choice $X_-^I=0$ and
supposing that the form fields $C_3$ and $C_6$ are not depend on
$x_+^I$ the MCS term of the full Lagrangian doesn't change the
solution\footnote{In general, $C_6$ may have a dependence
on $X_+^I$. If so, then our discussion is true only at leading
order. } $X_+^I=v^I$. So, one may rewrite
 \bea
 M^{IJK}=[X^I,X^J,X^K].
 \eea
 From the non-linear Lagrangian the
potential for the scalar is given by
 \bea\label{f8}
  -T_2STr\left
((det(S))^{1/4}\right)
\eea
 which can be expanded as \cite{Iengo:2008cq}
 \bea\label{f9}
(det(S))^{1/4}=1-\frac{1}{12T_2}M^{IJK}M^{IJK}+...
 \eea

 For the MCS term one has, at lowest order,  the following expression for the coupling of
$C_6$ flux to $M2$ branes world volume
 \bea\label{f10}
 i\lambda\mu_2\int STr\left(\frac{1}{216}\epsilon^{\mu\nu\lambda}M^{IJK}
C_{IJK\mu\nu\lambda}(x^{\mu},X^I)
+\frac{1}{72}\lambda\epsilon^{\mu\nu\lambda}M^{IJK}
C_{IJKL\nu\lambda}(x^{\mu},X^I)D_{\mu}X^L\right) \cr
 =i\lambda\mu_2\int d^3x STr\epsilon^{\mu\nu\lambda}
 M^{IJK}
 \left( \frac{1}{216}\left(C_{IJK\mu\nu\lambda}(x^{\mu})
 +\lambda
 X^L\partial_LC_{IJK\mu\nu\lambda}(x^{\mu})+...\right)\right)\hspace{1cm}\cr
 +i\lambda\mu_2\int d^3x STr\epsilon^{\mu\nu\lambda}
 M^{IJK}\left(\frac{1}{72}\left(\lambda C_{IJKL\nu\lambda}(x^{\mu})D_{\mu}X^L+...\right)\right),\hspace{3cm}
 \eea
 where in the second line we use the nonabelian Taylor expansion of
 $C_6$ form field.
 Now, consider that $M2$ branes are placed in a constant nontrivial $F_7=dC_6$ field strength as
\bea\label{f11}
 F_{\mu\nu\lambda IJKL}=\left\lbrace\matrix{-\frac{f}{3\lambda^2\mu_2}
 \epsilon_{\mu\nu\lambda}f_{IJKL}&{\rm for}\ I,J,K,L= \{3,4,5,6\}\cr 0&{\rm otherwise}\cr} \right.
 \eea
 where $f$ is a constant with dimension $(length)^2$.
 From the above assumptions, the first term in (\ref{f10}) equals
 to zero\footnote{This is true for constant background field or infinite dimensional
 algebra.}.
 By integrating by part of second term and noticing the fact that
 \bea
  F_{\mu\nu\lambda IJKL}=\frac{1}{48}\partial_{\mu}C_{\nu\lambda
  IJKL}-\frac{1}{36}\partial_{I}C_{\mu\nu\lambda JKL}
 \eea
  one obtains for the second and third terms
  \bea
  -i\lambda^2\mu_2\int d^3x
   Tr\left(\frac{1}{6}\epsilon^{\mu\nu\lambda}M^{IJK}X^LF_{\mu\nu\lambda
   IJKL}\right)
 \eea
 So, the potential for the scalars reads as
 \bea\label{f12}
 V(X)=-\frac{1}{12}STr
 \left(M^{IJK}M^{IJK}\right)-\frac{i\lambda^2\mu_2}{6}STr\left(
 \epsilon^{\mu\nu\lambda}M^{IJK}X^LF_{\mu\nu\lambda IJKL}\right)
 \eea
 Substituting (\ref{f11}), recalling $\delta X_+^I=0$ and demanding $\delta V(X)/\delta X^I=0$
 yields the equation
 \bea\label{f13}
 \left[M_{IJK},X^J,X_+^K\right]-if f^{IJKL}M_{JKL}=0,
 \eea
 which is the equation for extrema of $V(X)$.

 There is a trivial solution for this equation in which all fields
 are represented as diagonalized matrices as
 \bea\label{f14}
X^I=\pmatrix{x^I_1& 0& 0& \ddots\cr
                 0 &x^I_2& \ddots & 0\cr
          0&\ddots &\ddots& 0 \cr
         \ddots &0 &0 &x^I\cr} .
 \eea
 This is because of that from the commuting matrices (\ref{f14}) we
 have $M^{IJK}=0$ which solves the equation (\ref{f13}).

 As a nontrivial solution we use the following ansatz
 \bea\label{f17}
 M_{IJK}=2iRf_{IJKL}X^L,
 \eea
 which solves (\ref{f13}) if $R=\frac{f}{2}$.
 So, in this case we have the following equation
 \bea\label{f15}
 [X^I,X^J,X^K]=iff^{IJKL}X_L.
 \eea
 which has a simple solution
 \bea\label{f18}
  X^I=\sqrt{f}T^I,
 \eea
 Since the background field (\ref{f11}) breaks the $SO(8)$ symmetry to $SO(4)\times SO(4)$ one may naturally choose the structure constant to be $f_{IJKL}=\epsilon_{IJKL}$ \cite{Bagger:2006sk}.

 So, the solution (\ref{f18}) represents a fuzzy three-sphere \cite{Constable:2001kv} with radius
 $r^2=\Sigma(X^I)^2$.
 By defining $C=\Sigma(T^I)^2$ and noticing that since $C$ is the
 central of Lie-3 algebra (\ref{m1}) one can choose it to be a
 constant operator, we have
 \bea
 r^2=fC.
 \eea
 Evaluating the potential $V(X)$ using the solutions (\ref{f18}) we
 have
 \bea
 V(X)=-\frac{3}{2}f^3C
 \eea
 which is lower than that was obtained from the commuting solution
 (\ref{f14}), and means that the commuting solution is unstable
 and the system goes towards the formation of fuzzy $S^3$ configuration.

 \section{Summary} By placing a system of N $D_p$ branes in an
 external background form field causes that the system would has a
 vacuum in which is noncommutative an stable against of the
 commutative one \cite{Myers:1999ps}. Similarly, in M-theory, if a
 collections of membranes would be in an external $C_6$ form
 field the system has a vacuum in which membranes are polarized due
 to field strength effect and are formed into a fuzzy $S^3$ sphere. This
 can be interpreted as the formation of spherical branes ended on
 five-brane.

\end{document}